\documentclass[prd, 11pt,a4paper,twocolumn,groupedaddress,preprintnumbers,nofootinbib,floatfix]{revtex4-1}
\usepackage[top=3cm, bottom=2.1cm, left=2cm, right=2cm]{geometry}
\pdfoutput=1

\usepackage{comment}

\usepackage{graphicx}
\usepackage{url}
\usepackage[bookmarks, pagebackref=false]{hyperref}
\usepackage[usenames,dvipsnames]{xcolor}
\usepackage{dcolumn}
\usepackage{bm}
\usepackage{bbm}
\usepackage{amsmath,amssymb,amsfonts}
\usepackage{color}
\usepackage{hyperref}
\usepackage[Symbolsmallscale]{upgreek}
\usepackage{dsfont}
\usepackage[vcentermath]{youngtab}
\usepackage[all]{xy}
\usepackage{pstricks}
\usepackage{dsfont}%
\usepackage{mathtools}
\usepackage{placeins}
\usepackage{enumerate}
\usepackage{cases}
\usepackage{placeins}
\usepackage{xspace}
\usepackage{cancel} 
\usepackage{slashed}
\usepackage{subfigure}
\usepackage{natbib}
\hypersetup{
    bookmarks=true,         
    unicode=false,          
    pdftoolbar=true,        
    pdfmenubar=true,        
    pdffitwindow=false,     
    pdfstartview={FitH},    
    pdftitle={My title},    
    pdfauthor={Author},     
    pdfsubject={Subject},   
    pdfcreator={Creator},   
    pdfproducer={Producer}, 
    pdfkeywords={keyword1} {key2} {key3}, 
    pdfnewwindow=true,      
    colorlinks=true,       
    linkcolor=magenta,          
    citecolor=magenta,        
    filecolor=magenta,      
    urlcolor=cyan           
}
\usepackage{ulem}

\newcommand{\beq}{\begin{eqnarray}}
\newcommand{\eeq}{\end{eqnarray}}

\newcommand{\bmp}{\noindent\begin{minipage}{16cm}}
\newcommand{\emp}{\end{minipage}\vskip 7mm} 

    \newcommand{\ii}{\mathrm{i}}
    
    \newcommand{\ee}{\mathrm{e}}

    \newcommand{\SO}{\mathrm{SO}}

    \newcommand{\SUL}{\mathrm{SU}(2)_{\mathrm{L}}}

    \newcommand{\MS}{M_{\varphi}}
    
    \newcommand{\vw}{v_{\mathrm{w}}}
    \newcommand{\dl}{\delta_2}

    \newcommand{\myuk}{\alpha}

    \newcommand{\bee}{\begin{equation}}
        \newcommand{\eee}{\end{equation}}

\def\lsim{\mathrel{\rlap{\lower4pt\hbox{\hskip1pt$\sim$}}
    \raise1pt\hbox{$<$}}}                
\def\gsim{\mathrel{\rlap{\lower4pt\hbox{\hskip1pt$\sim$}}
    \raise1pt\hbox{$>$}}}                

\begin{document}

\title{Neutrino mass generation and leptogenesis via\\ pseudo-Nambu--Goldstone Higgs portal
}
\author{Tommi {\sc Alanne}}
\email{alanne@cp3-origins.net}
\affiliation{{CP}$^{ \bf 3}${-Origins},  University of Southern Denmark, Campusvej 55, DK-5230 Odense M, Denmark.}
\author{Aurora {\sc Meroni}}
\email{aurora.meroni@helsinki.fi}
\affiliation{Department of Physics, University of Helsinki,
\& Helsinki Institute of Physics, \\
                      P.O.Box 64, FI-00014 University of Helsinki, Finland.}
\author{Kimmo {\sc Tuominen}}
\email{kimmo.i.tuominen@helsinki.fi}
\affiliation{Department of Physics, University of Helsinki,
\& Helsinki Institute of Physics, \\
                      P.O.Box 64, FI-00014 University of Helsinki, Finland.}
\begin{abstract}
We consider an extension of the Standard Model with the global symmetry-breaking 
pattern SO(5)/SO(4), where the Higgs boson arises
as a pseudo-Nambu--Goldstone boson.
The scalar content of the theory consists of
a Standard-Model-like Higgs field and an extra real scalar field.
The flavour sector of the model is extended
by two right-handed neutrinos compatible with the observed light-neutrino
phenomenology, and we find that the correct vacuum alignment
determines the mass of the heavier neutrino eigenstate to be around 80~TeV.
The new singlet-scalar state  generates dynamically a Majorana mass
term for the heavy neutrino states.
We show how the model leads to the correct baryon asymmetry
of the universe via leptogenesis in the case of two degenerate
or hierarchical heavy neutrinos.\\
\\
{\footnotesize \it Preprint: CP$^3$-Origins-2017-024 DNRF90 \& HIP-2017-12/TH}
\end{abstract}
\maketitle
\newpage

\section{Introduction}
\label{sec:intro}

After the discovery of the Higgs boson
and thereby the verification of the Standard-Model (SM) like electroweak symmetry-breaking (EWSB) pattern to very high accuracy,
the neutrino sector provides one of the most prominent
sources for beyond-SM (bSM) phenomenology.
In addition to not explaining the neutrino mass and mixing patterns, the SM
does not contain fields which would act
as dark matter (DM), and it does not allow for dynamical explanation of baryon asymmetry
of the Universe (BAU).

These issues provide motivation to explore extensions of the SM even if the
present collider experiments have not revealed any direct signals
of bSM physics. The absence of direct signals must be interpreted to imply that the new physics occurs either at scales beyond the current experimental reach and/or is very weakly coupled with the SM.

Here we contemplate the idea that the EWSB and neutrino mass generation be connected via a minimal
extension of the SM scalar sector, and the neutrino sector, in turn, could mend the SM shortcomings in explaining the BAU and DM abundance.
In this paper, we focus on the details of BAU and comment on the possible routes
towards DM model building.

An attractive  explanation of the Majorana nature of massive neutrinos is
provided by the See-Saw mechanism~\cite{Minkowski:1977sc,GellMannRamondSlansky,Yanagida,Mohapatra:1979ia,Schechter:1980gr,Schechter:1981cv}, which not only gives an
explanation of the smallness of neutrino masses via heavier fermionic
singlets, but also gives an explanation to the
observed BAU through leptogenesis~\cite{Kuzmin:1985mm,Fukugita:1986hr}.

Connecting the neutrino mass generation to EWSB and the details of flavour physics
likely require a larger scalar sector than the SM one.
The compatibility with the spectrum observed at the LHC can be achieved
in models predicting the Higgs as a pseudo-Nambu--Goldstone boson (pNGB);
see e.g. Ref.~\cite{Schmaltz:2005ky} for a review of this type of models.
The Higgs sector can be either elementary or composite.
While the composite case is attractive as it allows to address
the hierarchy problem, it lacks simple dynamics to produce the
SM-fermion masses.
The elementary case, on the other hand,
provides a calculable framework to
assess the observed symmetry-breaking pattern and
low-energy spectrum~\cite{Coleman:1973jx}, and
facilitates an effective description of flavour physics in terms of
Yukawa interactions~\cite{Alanne:2016mmn} as in the SM. We will follow the latter route in this paper.

We will investigate the  $\SO(5)\rightarrow\SO(4)$ pattern of global symmetry breaking as a
renormalizable field theory
simultaneously protecting the Higgs mass and
reproducing a Majorana neutrino mass term.
The  coset of the global symmetry breaking pattern, $\SO(5)/\SO(4)$,
allows the presence of  a SM scalar doublet using the fundamental
representation for the scalar sector.
Explicitly, the fundamental of $\SO(5)$ decomposes as
$\mathbf{5}= \mathbf{1} \oplus \mathbf{(2,2)}$, and therefore a scalar singlet
appears in the model. The SM extension  featuring  a scalar
singlet (also dubbed SSM) is a common renormalizable augmentation of the Higgs
sector  and has been scrutinized in the
literature; see e.g. Refs.~\cite{Pruna:2013bma, Englert:2011yb,
Batell:2011pz, Buttazzo:2015bka, Robens:2015gla,
Robens:2016xkb,McDonald:1993ex,Cline:2012hg,Alanne:2014bra}.
Our model differs essentially from these models, since
in our case the singlet exists as a part of larger multiplet due to
a larger global symmetry.

We extend the SO(5)/SO(4) model by adding two right-handed (RH) neutrinos whose
interaction and mixing patterns are chosen to conform with experimental
results. We will find
that the presence of these fields will orient the vacuum so that proper EWSB
pattern ensues.
This, together with properties of the vacuum, provides non-trivial constraints on the model parameters.
Finally, the new matter fields will allow observed BAU to be generated via leptogenesis.
The details of the mechanism are sensitive to whether
the masses of the RH neutrinos are hierarchical or almost degenerate.

We find that observed BAU, together with correct pattern of EW symmetry breaking
with spectrum compatible with observations, is achieved in the following scenario: the SO(5) symmetry breaks at scale of order of $v\sim 10^4$ TeV, and the heaviest neutrino mass eigenstate has mass around 80 TeV. The Higgs boson with mass 125 GeV is mostly Goldstone-like state, and the heavier singlet state in the SO(5) multiplet is assumed to be lighter than the heaviest RH neutrino. This forces the scalar self-coupling to be tiny, of order $10^{-8}$. Cosequently, the heavier scalar state has mass $m^2\sim\lambda v^2\sim {\cal{O}}(1\,{\rm{TeV}}^2)$, and the trilinear Higgs coupling will be of the order of ${\cal{O}}(10^{-4}\,{\rm{TeV}})$.

The paper is organized as follows: In Sec.~\ref{sec:model} we outline the
details of the model. In Sec.~\ref{sec:vev} we analyse
vacuum structure, and how this constrains the
parameter space of the model.
In Sec.~\ref{sec:leptogenesis} we discuss how the observed BAU
can be produced in this model. We analyse quantitatively the cases of degenerate or hierarchical neutrinos.
In Sec. \ref{sec:checkout} we present our conclusions and outlook for further work.

\section{The Model}
\label{sec:model}

In this work we consider the minimal extension of the SM scalar sector
incorporating an elementary pNGB Higgs.
We adopt a tree-level scalar potential that features a global symmetry
breaking pattern $\SO(5)\rightarrow\SO(4)$.
The  scalar degrees of freedom are conveniently parameterized by a multiplet
$\Sigma$ describing a linear $\sigma$-model based on the coset $\SO(5)/\SO(4)$.
We adopt the following notation
\bee\Sigma= (\sigma,\Pi_1,\Pi_2,\Pi_3,\Pi_4 ),\eee
where $\sigma$ is a massive scalar degree of freedom and $\Pi_i$, $i=1,\dots, 4$ are the four Nambu--Goldstone (NGB) fields associated with the broken generators of SO(5). The general SO(5)-symmetric potential is given by
\bee V= \frac{1}{2}m_\Sigma^2\Sigma^T \Sigma + \frac{\lambda}{4!} (\Sigma^T \Sigma)^2. \label{eq:V0}\eee

The electroweak (EW) gauge group, $\SUL\times \mathrm{U}(1)_Y$,
is embedded in $\SO(5)$
such
that the three broken generators of $\SUL$ are those
associated with the three NGB's $\Pi_i$, $i=1,\dots, 3$.
The field $\Sigma$ contains an EW Higgs doublet, $H$, and another
real singlet, $\varphi$, which are the EW interaction eigenstates.
The relevant physical mass eigenstates will in general
be mixtures of the neutral component of $H$ and $\varphi$
such that the Higgs is mostly the NGB-like state, while the
heavier singlet neutral scalar is mostly the $\sigma$-like eigenstate.

Explicitly,
\bee
H=\frac{1}{\sqrt{2}}\begin{pmatrix} \Pi_1+\ii \Pi_2 \\ h+\ii \Pi_3\end{pmatrix},
\eee
and in the unitary gauge we can write the potential of Eq.~\eqref{eq:V0} in terms of the neutral components of the
EW eigenstates,
$h=\cos\theta\,\Pi_4+\sin\theta\,\sigma$ and $\varphi=-\sin\theta \,\Pi_4 +\cos\theta\, \sigma$,
\begin{equation}
   \label{V0}
   \begin{split}
 V_0=&\,\frac{m_H^2}{2}h^2+\frac{m_{\varphi}^2}{2}\varphi^2\\
     &+ \frac{1}{4!}\lambda_Hh^4+\frac{\lambda_\varphi}{4!} \varphi^4+ \frac{\lambda_{H\varphi }}{12} h^2\, \varphi^2,
   \end{split}
\end{equation}
with  $m_H^2(\mu_0)=m_{\varphi}^2(\mu_0)\equiv m_{\Sigma}^2$,
$\lambda_H(\mu_0)=\lambda_{\varphi}(\mu_0)=
\lambda_{H\varphi}(\mu_0)\equiv\lambda$,
where the scale $\mu_0$ is of the order of the symmetry-breaking scale and
is determined by renormalization conditions\footnote{Note that
one can relate the masses and couplings of different field components to the
corresponding parameters in the SO(5)-invariant Lagrangian only at the scale
where SO(5) becomes effectively restored as the couplings will run
differently.}.

The radiative symmetry-breaking dynamics implies:
\begin{equation}\label{eq:vH}
    \langle h\rangle=v\sin\theta\equiv v_{\rm{w}}=246\, \mathrm{GeV},
\end{equation}
and
\begin{equation}\label{eq:vphi}
    \langle \varphi\rangle=v\cos\theta.
\end{equation}
Radiative corrections single out a value for the angle $\theta$, and the large hierarchy $v_{\rm{w}}\ll v$ is reflected in $\sin\theta\ll 1$.

Furthermore, we introduce  two RH Majorana neutrinos.
We couple them via Yukawa interactions
to the EW-singlet scalar state,  $\varphi$, such that the
Majorana masses for the RH neutrinos are induced via the symmetry breaking.
However, a generic Majorana mass term for RH neutrinos is allowed by gauge
interactions, and hence the most generic Lagrangian we can write is:
\begin{equation}
    \label{eq:nuYuk2}
    \begin{split}
    -\mathcal{L}^{\nu}=\, &  Y_{ij}\overline{N_{Ri}} (L \tilde H)_j +\frac{1}{2}\overline{(N_R)^c}_i\mathcal{M}_{ij} N_{Rj}\\
    & +\,\frac{1}{2}\myuk_{ij}\overline{(N_R)^c}_i N_{Rj}\,\varphi +{\textrm{h.c.}}\\
     =\, &  M_D\overline{N_R}\nu_L +\frac{1}{2}\overline{(N_R)^c}
     M_{N} N_{R} +{\textrm{h.c.}}
    \end{split}
\end{equation}
where
\bee
    M_D= \vw Y \quad \mbox{and}\quad M_{N}=\mathcal{M} +\langle\varphi\rangle\myuk.
    \label{eq:masses}
\eee
We note that $M_N$ must be symmetric due to the Majorana condition $(N_R)^c= C\overline{N_R}^T$.
The Lagrangian in  Eq.~\eqref{eq:nuYuk2} is a realisation of the type I See-Saw.
It is always possible to diagonalize $M_N$, i.e.
there exists a matrix
$V$ such that
\bee \mathcal{M} +\langle\varphi\rangle\myuk= V D_N V^T,
\eee
with the heavy-neutrino mass matrix defined as
$(D_N)_{ij}\equiv m_{Ni}\delta_{ij}$.

\section{Vacuum alignment analysis}
\label{sec:vev}

In the following, we assume a simplified structure of
the bare Majorana mass matrix ${\cal M}={\rm{diag}}(M_0,M_0)$, and
consider two specific textures for $\alpha$: diagonal and almost democratic,
\bee
\alpha_0\begin{pmatrix} 1-\epsilon & 0 \\ 0 & 1\end{pmatrix},\quad
\alpha_0\begin{pmatrix} 1 & 1-\epsilon \\ 1-\epsilon & 1\end{pmatrix},
\label{eq:twoCases}
\eee
where $\epsilon\ll 1$ is introduced to provide small splitting of masses in the diagonal case and
to provide a non-zero mass contribution for the lightest eigenstate in the democratic case.
In these two cases, the mass of the heavier
mass eigenstate
is, respectively, $M_0+|\alpha_0|v\cos\theta$
and $M_0+2|\alpha_0|v\cos\theta$,
up to corrections of ${\cal O}(\epsilon)$.
Next we consider the constraints arising from the vacuum alignment
on the parameters $M_0$ and $\alpha_0$. We will find that vacuum alignment
is mainly determined by $\alpha_0$, while the correct value of the Higgs boson
mass imposes an upper limit for $M_0$.

As shown in Ref.~\cite{Alanne:2016mmn},  in the absence of the RH-neutrino sector,
the EW gauge and top corrections prefer an unbroken EW symmetry, i.e. the
value $\theta=0$. Switching on the coupling $\alpha_{ij}$ between
the RH neutrinos and the singlet $\varphi$, the picture changes, implying a
non-trivial alignment angle, $\theta$.
This is due to the different dependence on the vacuum alignment angle of the RH
neutrinos and the SM sector, as can be seen from Eqs.~\eqref{eq:vH}
and~\eqref{eq:vphi}.

To illustrate the main features of the vacuum structure in the model under
study, we first consider the simplest case setting $M_0=0$ in
Eq.~\eqref{eq:masses} and $\epsilon=0$ in Eq.~\eqref{eq:twoCases}.
For the democratic $\alpha$-matrix, this corresponds to one
heavy neutrino since
there are two eigenvalues, one of which is zero. For the diagonal
$\alpha$-matrix, this corresponds to two degenerate heavy neutrinos.

With the choice $M_0=0$, one can obtain simple analytic expressions.
Moreover, the result on the alignment angle $\theta$ is essentially
not affected by non-zero values of $M_0$. This is due to the fact that
allowing non-zero $M_0$ does not alter the $\theta$-dependence of the
neutrino contribution to the one-loop effective potential.

To arrive at a simple analytic expression for the non-trivial solution for the
vacuum alignment angle, we also ignore the one-loop corrections
from the scalar sector. In the numerical analysis, we include all these
corrections and show that the results for the alignment do not essentially
change from the simplified results.
In this simplified limit, the effective potential up to the one-loop level can be written as
\begin{equation}
    \label{eq:}
    V_{\mathrm{eff}}=V_0+V_1^{\mathrm{SM}}+V_1^{\nu},
\end{equation}
where following the notations of Ref.~\cite{Alanne:2016mmn}, the one-loop contributions in the $\sigma$-background and in the $\overline{\mathrm{MS}}$
scheme can be written as
\begin{equation}
    \label{eq:}
    \begin{split}
	V_1^{\mathrm{SM}}=&\frac{3g^4\sigma^4\sin^4\theta}{64\pi^2}\\
	&\cdot\left(A_{\mathrm{SM}}\log\frac{g^2\sigma^2\sin^2\theta}{\mu_0^2}
	    +B_{\mathrm{SM}}\right),\\
	V_1^{\nu}=&\frac{|\myuk_0|^4}{32\pi^2}\sigma^4\cos^4\theta\\
	&\cdot\left(A_{\nu}\log\frac{|\myuk_0|^2\sigma^2\cos^2\theta}{\mu_0^2}
	    +B_{\nu}\right)
    \end{split}
\end{equation}
where we define
    \begin{align}
	\label{eq:}
	    A_{\mathrm{SM}}=&\frac{1}{16} \left[\left(\frac{g^2+g^{\prime\,2}}{g^2}\right)^2+2\right]-\frac{y_t^4}{g^4},\nonumber\\
	    B_{\mathrm{SM}}=& \frac{1}{16} \left[\frac{8y_t^4}{g^4}\left(3-\log\frac{y_t^4}{4g^4}\right)-\frac{5}{3}-\log 16\right.
		\nonumber\\
	    &\left.+\left(\frac{g^2+g^{\prime\,2}}{g^2}\right)^2\left(\log\frac{g^2+g^{\prime\,2}}{4g^2} -\frac{5}{6}\right)\right],
		\nonumber\\
	    A_{\nu}^{\mathrm{diag}}=&-2,\quad A_{\nu}^{\mathrm{democ}}=-16,\\
	    B_{\nu}^{\mathrm{diag}}=&\,3,\quad B_{\nu}^{\mathrm{democ}}=-16\left(2\log 2-\frac{3}{2}\right).\nonumber
    \end{align}

We choose the renormalization scale, $\mu_0$, such that the tree-level vacuum expectation value, $\langle\sigma\rangle=v$, is not
changed by the one-loop corrections and determine the preferred value of
the vacuum alignment by minimizing the effective potential with respect to $\theta$. This yields the following non-trivial solution
for $\theta$:
\begin{equation}
    \label{eq:}
    \begin{split}
	\tan^2\theta=&-\frac{2A_{\nu}|\myuk_0|^4}{3A_{\mathrm{SM}}g^4}\\
	=&\frac{512 |\myuk_0|^4}{3 \left(16y_t^4-3 g^4-2 g^2 g^{\prime\,2}-g^{\prime\,4}\right)}.
    \end{split}
\end{equation}
We assume the SM-like RG running of the gauge and top-Yukawa couplings up to the renormalisation scale, $\mu_0$.

In this simplified limit,
the $\overline{\mathrm{MS}}$ mass of the Higgs can be written as
\begin{widetext}
    \begin{equation}
	\label{eq:}
	\begin{split}
	    (m_h^{\overline{\mathrm{MS}}})^2=\frac{3g^4\vw^2}{16\pi^2 A_{\nu}\left(2A_{\nu}|\myuk_0|^4-3A_{\mathrm{SM}}g^4\right)}&\left[
		\vphantom{\frac{1}{2}}
		-3A_{\mathrm{SM}}g^4\left(2A_{\mathrm{SM}}A_{\nu}-3A_{\mathrm{SM}}B_{\nu}+3A_{\nu}B_{\mathrm{SM}}\right)\right.\\
	    &+2A_{\nu}|\myuk_0|^4\left(2A_{\mathrm{SM}}A_{\nu}+3A_{\mathrm{SM}}B_{\nu}-3A_{\nu}B_{\mathrm{SM}}\right)\\
	    &\left.+3A_{\mathrm{SM}}A_{\nu}\left(3A_{\mathrm{SM}}g^4+2A_{\nu}|\myuk_0|^4\right)\log\left(
		-\frac{3A_{\mathrm{SM}}g^2}{2A_{\nu}|\myuk_0|^2}\right)\right].
	\end{split}
    \end{equation}
\end{widetext}
Once we have the expression for the $\overline{\mathrm{MS}}$ mass, we calculate the physical mass
following Refs.~\cite{Casas:1994us,Gonderinger:2009jp}.

\begin{figure}
    \begin{center}
	\includegraphics[width=0.48\textwidth]{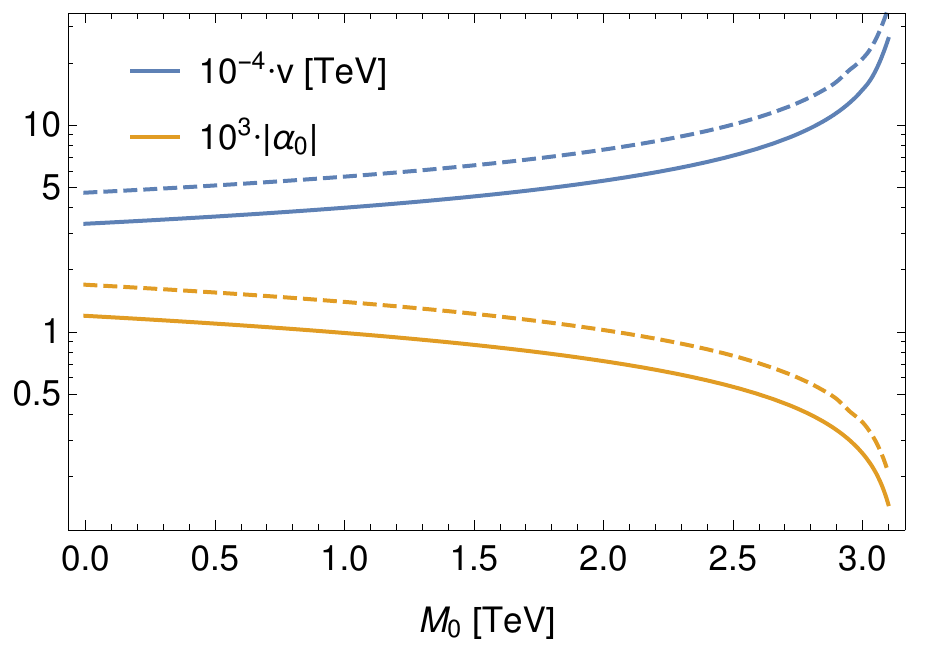}
    \end{center}
    \caption{The values of the vacuum expectation value and the (absolute value of the) Majorana Yukawa coupling, $|\alpha_0|$, as 
    a function of $M_0$,
    when the correct Higgs mass and the vacuum solution are imposed.
    Solid (dashed) lines correspond to democratic (diagonal) $\alpha$-matrices in Eq.\eqref{eq:twoCases}.}
    \label{fig:M0lim}
\end{figure}

\begin{figure}
    \begin{center}
	\includegraphics[width=0.48\textwidth]{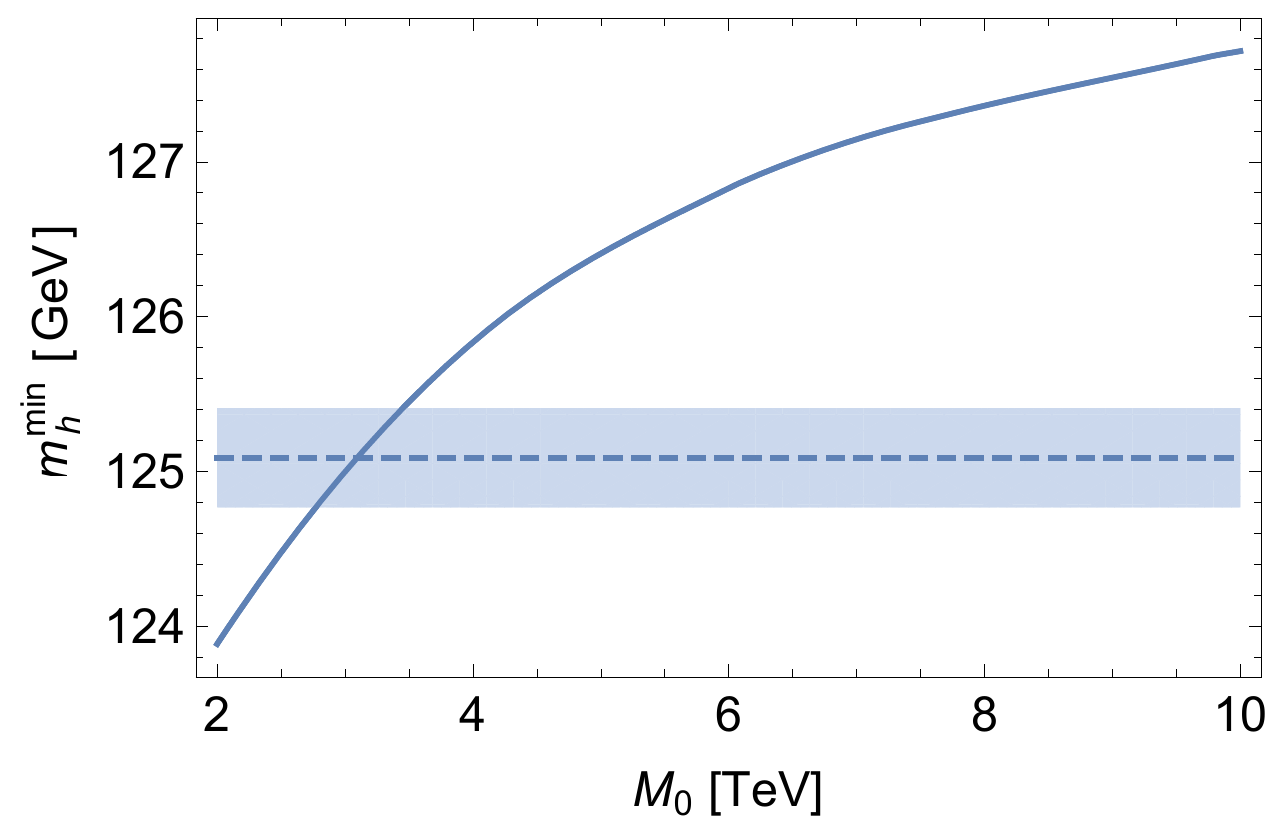}
    \end{center}
    \caption{The minimum allowed physical Higgs mass value as a function of $M_0$
    when the vacuum solution is imposed. The figure corresponds to the democratic texture in Eq.~\eqref{eq:twoCases}. The result of the diagonal case is practically indistinguishable.}
    \label{fig:M0mhlim}
\end{figure}

Next, we turn to the full numerical analysis including the one-loop corrections from the scalar sector and adding
a non-zero Majorana mass parameter, $M_0$.
The values for the symmetry-breaking scale and the absolute value of the Majorana-neutrino Yukawa
coupling, $|\myuk_0|$, for a stable minimum solution producing the
correct Higgs mass are given in Fig.~\ref{fig:M0lim}.
We find that
increasing the hard Majorana mass parameter $M_0$, while requiring a
stable minimum with respect to the angle $\theta$, constrains the
minimum allowed physical mass of the Higgs.
We show this dependence in Fig.~\ref{fig:M0mhlim}.

\begin{figure}
	\begin{center}
	    \includegraphics[width=0.45\textwidth]{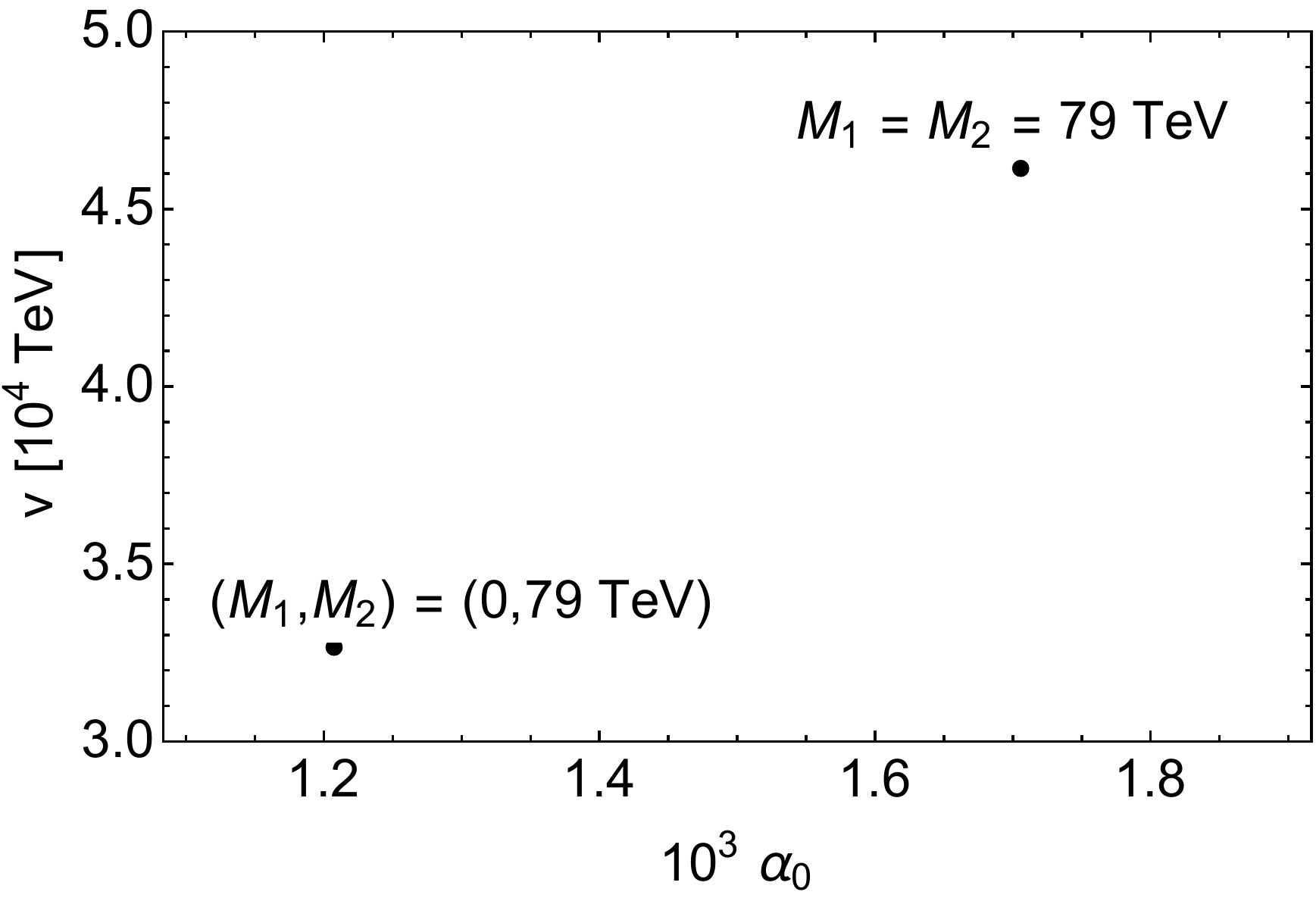}
	\end{center}
	\caption{The values for the symmetry-breaking scale as a function of the Majorana neutrino Yukawa coupling for one and two heavy RH neutrinos with
	    identical couplings. The points correspond to values
      at $M_0=0$ in Fig.~\ref{fig:M0lim}.}
	\label{fig:Nnudep2}
   \end{figure}

The democratic or diagonal $\alpha$-matrix implies two hierarchically separated
or two degenerate mass eigenvalues and, therefore, corresponds respectively to
one or two heavy RH flavours. In Fig.~\ref{fig:Nnudep2} we show how the value
of the symmetry-breaking scale and the value of the
Majorana Yukawa coupling depend on
these two possibilities for the RH-neutrino mass textures. We have fixed
$m_h^{\overline{\rm{MS}}}=128$\,GeV to account for the correct
physical Higgs mass.
We note that in both cases the value of the heaviest neutrino turns out to
be around 80 TeV.

\section{See-Saw mechanism and Leptogenesis}
\label{sec:leptogenesis}

Baryogenesis via leptogenesis~\cite{Fukugita:1986hr} is a simple mechanism to explain the
baryon number asymmetry of the
Universe (BAU). The asymmetry is usually expressed in terms of the net baryon-to-photon ratio \cite{Ade:2015xua}
\bee \eta_B=\frac{n_B- n_{\bar B}}{n_\gamma} \simeq (6.08 \pm 0.04) \times 10^{-10}.\eee

The effect of non-perturbative $(B+L)$-violating sphaleron processes can convert
a lepton asymmetry into a
baryon asymmetry~\cite{Kuzmin:1985mm,Fukugita:1986hr}.
This scenario can well be implemented within the See-Saw (type I) mechanism~\cite{Minkowski:1977sc,GellMannRamondSlansky,Yanagida,Mohapatra:1979ia}: in the
thermal leptogenesis scenario, after the inflation period,
the heavy RH neutrinos are produced by
thermal scatterings, and subsequently they can decay
out-of-equilibrium producing both lepton and CP number violation,
therefore satisfying all of the Sakharov's conditions (see Ref.~\cite{Riotto:1999yt}
for a review).

In this Section, we study the different possibilities to produce
the observed amount of the BAU within
the model presented in the previous section.
We restrict ourselves
to the minimal RH-neutrino sector with two heavy flavours to illustrate the main features of generating BAU.
We note, however, that extending the neutrino sector with a third
RH neutrino would allow for a DM candidate alike the one considered
e.g. in Refs.~\cite{Asaka:2005an,Asaka:2005pn,DiBari:2016guw}.
Furthermore, the addition of a third, lighter RH
neutrino would not affect the vacuum structure, since this is predominantly determined by the heaviest state.
We leave, however, the study of DM phenomenology to future work, and concentrate here on details of how to generate the BAU.

Interestingly the two simple forms of $\alpha$ given
in Eq.~\eqref{eq:twoCases}
imply different possibilities for the production of the baryon asymmetry
in this model: the diagonal
form will allow for resonant leptogenesis, while the democratic one will allow
for non-resonant standard scenario.

We express the neutrino Dirac-Yukawa couplings through the Casas--Ibarra parameterization~\cite{Casas:2001sr}.
We notice that this parameterization is valid also in presence of
additional contributions to the bare
Majorana-neutrino mass  term.
Indeed, in a generic basis for the heavy Majorana-neutrino mass matrix, $M_N$, we write:
\begin{eqnarray}
	M_D&=&i U_{\mathrm{PMNS}}^*\sqrt{D_\nu}\, \Omega\sqrt{D_N} V^\dagger,
\end{eqnarray}
with the diagonal heavy and light neutrino mass matrices defined
respectively as $(D_N)_{ij}\equiv m_{Ni}\delta_{ij}$  and
$(D_\nu)_{ij}\equiv m_i \delta_{ij}$,  with $m_i< 1$ eV and $i=1,2,3$.

The matrix   $U_{\mathrm{PMNS}}$ is the
Pontecorvo--Maki--Nakagawa--Sakata matrix describing leptonic interactions, $\Omega$
is a general $3\times3$
complex matrix that can be parameterized as
$\Omega=R(\theta_{23})R(\theta_{13})R(\theta_{12})$ with $\theta_{ij}$
complex, and
$V$ is the matrix diagonalizing the
RH Majorana mass matrix. Similarly to the standard type I See-Saw scenario,
one can go to the basis where the
the RH Majorana mass matrix is diagonal, i.e. $V$ is set to unity.

In the analysis that follows, we will restrict ourselves to the minimal case
of two heavy RH neutrinos,
\begin{equation}
    \label{eq:}
    M_N=\mathrm{diag}(M_1, M_2),
\end{equation}
and parameterize two different regimes in terms of $\delta\equiv (M_2-M_1)/M_1$:
\begin{itemize}
    \item degenerate masses for $\delta\ll 1$ and
    \item hierarchical spectrum for  $\delta\gg 1$,
\end{itemize}
corresponding to our choice of a diagonal or a democratic $\alpha$-matrix, respectively.
Producing the correct EW spectrum and the Higgs mass fixes the mass of
the heaviest RH neutrino to be not heavier than about 80 TeV.
Therefore, the
simplest hierarchical leptogenesis scenario
is not possible due to the
Davidson--Ibarra bound~~\cite{Davidson:2002qv} on the lightest RH neutrino
mass. A possible solution to circumvent this bound, and achieve a viable
low-energy leptogenesis scenario is to consider nearly degenerate
RH-neutrino masses, the so-called resonant leptogenesis~\cite{Pilaftsis:1997jf, Pilaftsis:2003gt}.
We will consider this scenario in Sec.~\ref{sec:LG-1}.

An alternative low-energy scenario considered in Ref.~\cite{Dall:2014nma}
is the case where the couplings between the RH neutrinos and the singlet scalar
provide an additional source of CP violation, and allow to avoid the Davidson--Ibarra bound.
In this case both the bare Majorana mass and the off-diagonal mass term,
$\alpha_{ij}$, $i\neq j$, are needed to produce a successful baryogenesis scenario.
We consider a minimal extension of the model to allow for this leptogenesis
scenario in Sec.~\ref{sec:LG-2}.

    \subsection{Scenario I: Resonant leptogenesis}
    \label{sec:LG-1}

    Here we consider the case where the two RH neutrinos are
    quasi-degenerate, and the oscillations between different neutrino flavours
    provide a sufficient enhancement of
    CP violation.  This scenario is dubbed as resonant leptogenesis~\cite{Pilaftsis:1997jf, Pilaftsis:2003gt}, and it has been
    extensively studied
    in connection to the  Dirac and/or Majorana CP-violating phases in the PMNS neutrino mixing matrix~\cite{Pascoli:2006ci}. For recent developments, see e.g. Refs.~\cite{Garny:2011hg,Dev:2014wsa,Drewes:2016gmt}.
    This scenario is particularly interesting since  it can directly relate the low-energy CP violation in the lepton sector
    to the BAU.

    To estimate the asymmetry, we follow Refs.~\cite{Pascoli:2006ie,Pascoli:2006ci,DiBari:2016guw}, and
    the corresponding details about the resonant leptogenesis scenario can be found in Appendix~\ref{app:resonant}.
    We consider two heavy RH neutrinos with masses $M_1$ and $M_2$ which,
    in the notation of Eq.~\eqref{eq:nuYuk2}, satisfy
    \begin{equation}
	\label{eq:limYuk}
	\delta\equiv(M_2-M_1)/M_1 \gg \frac{(Y^{\dagger}Y)_{12}}{16\pi^2}.
    \end{equation}

    We assume that the baryon-to-photon ratio at recombination satisfies $\eta_B\simeq 0.01 N_{B-L}^{f}$,
    where $N^f_{B-L}$ is explicitly given in Eq.~\eqref{eq:nbminusl}.
    The proportionality constant contains the sphaleron conversion factor, $a_{\rm{sph}}$, which for SM particle content is $28/79\sim 1/3$, and the
    dilution due to the photon production between the leptogenesis scale and the scale of recombination, $1/f=86/2387\sim 1/30$ assuming the standard
    isentropic expansion~\cite{Buchmuller:2004tu}.

    The end result of the analysis for the normal hierarchy of the active neutrinos is
    \begin{equation}
	\label{eq:resonantresult}
	\delta
	\simeq 0.8\times 10^{-7}\left(\frac{f(m_{\nu},\Omega)}{f_{\max}}\right)\left(\frac{M_1}{10^4\,\mathrm{GeV}}\right),
    \end{equation}
    where $f_{\max}\simeq 0.005$, and the function $f$, defined explicitly in Appendix \ref{app:resonant},
    incorporates the dependence on the light-neutrino masses, $m_\nu$, and mixings via $\Omega$. With RH neutrinos at scale
    $M_1\sim 80$ TeV, the degeneracy of the order
    of $\delta\lesssim 10^{-7}$ is expected for a viable leptogenesis scenario.

    Hence, this scenario requires considerable fine tuning in the mass spectrum, but nevertheless provides
    a viable scenario to produce the observed BAU.

    \subsection{Scenario II: Hierarchical masses}
    \label{sec:LG-2}

    Let us now consider leptogenesis in the case where the RH neutrinos of the model have hierarchical masses,
    i.e. $\delta=(M_2-M_1)/M_1\gg 1$.
    Our model contains the same degrees of freedom as the one considered in detail in Ref.~\cite{Dall:2014nma}.
    We therefore follow their prescription to implement the two-stage Boltzmann equations taking the interactions
    of the singlet scalar into account. The relevant processes for the Boltzmann equations
    are ${\cal O}(Y^2)$, ${\cal O}(Y^4)$, and ${\cal O}(Y^2\alpha^2)$ corresponding to $\Delta L=1$ decays, $\Delta L=2$ scatterings,
    and $\Delta L=1$ scatterings, respectively.

    The additional CP violation due to the extra scalar is generated by $N_2\rightarrow LH$ decays via
    $N_1N_2\varphi$ interactions. The lighter neutrino, $N_1$, remains in equilibrium and can mediate potentially dangerous washout processes.
    However, if $N_1$ is weakly-enough coupled, then this washout effect is suppressed, and does not disturb the enhancement of the CP violation~\cite{Dall:2014nma}.

    The coupled equations for the abundances of the two neutrino
    flavours, $Y_1$ and $Y_2$, and the lepton asymmetry $Y_{L-\bar{L}}$ are
     \begin{widetext}
	\begin{align}
	    \label{eq:BEs}
	    z_1\frac{\partial Y_{L-\bar{L}}}{\partial z_1}=&\epsilon_1D_1\left(\frac{Y_1}{Y_1^{\mathrm{eq}}}-1\right)
		+\epsilon_2D_2\left(\frac{Y_2}{Y_2^{\mathrm{eq}}}-1\right)- Y_{L-\bar{L}}\left(W_{1D_1}+W_{S_1}+W_{1D_2}+W_{S_2}\right),\\
	    z_1\frac{\partial Y_1}{\partial z_1}=&-\left(\frac{Y_1}{Y_1^{\mathrm{eq}}}-1\right)\left(D_1+D_{21}+S_1\right)
		+\left(\frac{Y_2}{Y_2^{\mathrm{eq}}}-1\right)D_{21}\\
	    &-\left(\frac{Y_1Y_2}{Y_1^{\mathrm{eq}}Y_2^{\mathrm{eq}}}-1\right)S_{N_1N_2\rightarrow HH}
		-\left(\frac{Y_1^2}{Y_1^{\mathrm{eq}\,2}}-1\right)S_{N_1N_1\rightarrow HH},\\
	    z_1\frac{\partial Y_2}{\partial z_1}=&-\left(\frac{Y_2}{Y_2^{\mathrm{eq}}}-1\right)\left(D_2+D_{21}+S_2\right)
		+\left(\frac{Y_1}{Y_1^{\mathrm{eq}}}-1\right)D_{21}\\
	    &-\left(\frac{Y_1Y_2}{Y_1^{\mathrm{eq}}Y_2^{\mathrm{eq}}}-1\right)S_{N_1N_2\rightarrow HH}
		-\left(\frac{Y_2^2}{Y_2^{\mathrm{eq}\,2}}-1\right)S_{N_2N_2\rightarrow HH},
	\end{align}
    \end{widetext}
    where $Y_i=n_i/n_{\gamma}^{\rm{eq}}$\,, $z_i\equiv M_i/T$\,, $z_2=z_1M_2/M_1$, and
    \begin{equation}
	\label{eq:}
	 Y_i^{\mathrm{eq}}\equiv\frac{3}{8}z_i^2K_2(z_i).
    \end{equation}
    Here $K_2(z)$ is the modified Bessel function of the second kind.
    The CP violation is parameterized by
    \begin{equation}
	\label{eq:}
	\epsilon_i\equiv\frac{\Gamma(N_i\rightarrow LH)-\Gamma(N_i\rightarrow \bar{L}\bar{H})}{\Gamma(N_i\rightarrow LH)+\Gamma(N_i\rightarrow \bar{L}\bar{H})},
    \end{equation}
    and the functions denoted by $\Gamma$, $D$, $S$, and $W$ correspond to different decay, scattering and washout processes. The functions
    $D$, $S$, and $W$ are defined in Appendix~\ref{app:def}, while the decay rates, $\Gamma$, are defined in Appendix~\ref{app:decayW}.

    We can now directly proceed to present the numerical results we obtain for
    our model. The scalar sector has three parameters, $m_\Sigma$, $\lambda$ and the angle, $\theta$, while the
    leptonic sector depends on $\alpha_0$, $\epsilon$, $M_0$ and the Yukawa matrix, $Y$.
    The constraints on
    vacuum alignment and correct mass for the Higgs boson leave $M_0$, $\epsilon$, $Y$, and $\lambda$ as free parameters.
    The Yukawa matrix, $Y$,  enters through the decay rates $\Gamma_i= \Gamma(N_i\rightarrow LH)+\Gamma(N_i\rightarrow \bar{L}\bar{H})$
    and $\epsilon$ through $\Gamma_{21}=\Gamma(N_2\rightarrow N_1\varphi)$ given explicitly in Eq.~\eqref{eq:gammaij}. A convenient parametrization
    is given by considering the equilibrium parameters  $K_i$, $i=1,2$, and $K_{12}$,
    \begin{equation}
	\label{eq:}
	K_i\equiv\frac{\Gamma_i}{H_i}=\frac{\tilde{m}_i}{m_{\star}}, \quad
	K_{21}\equiv\frac{\Gamma_{21}}{H_2},
    \end{equation}
    and we define
    \begin{equation}
	\label{eq:}
	\tilde{m}_i\equiv\frac{(YY^{\dagger})_{ii}\vw^2}{2M_i},\quad m_{\star}\equiv 4\pi\vw^2\sqrt{\frac{8\pi^3g_{*}}{90M_{\mathrm{P}}^2}},
    \end{equation}
    and the Hubble rate is given by:
    \begin{equation}
	\label{eq:}
	H(T)\equiv T^2\sqrt{\frac{8\pi^3g_{*}}{90M_{\mathrm{P}}^2}},\quad H_i\equiv H(T=M_i).
    \end{equation}

\begin{figure}
 	\begin{center}
 	    \includegraphics[width=0.48\textwidth]{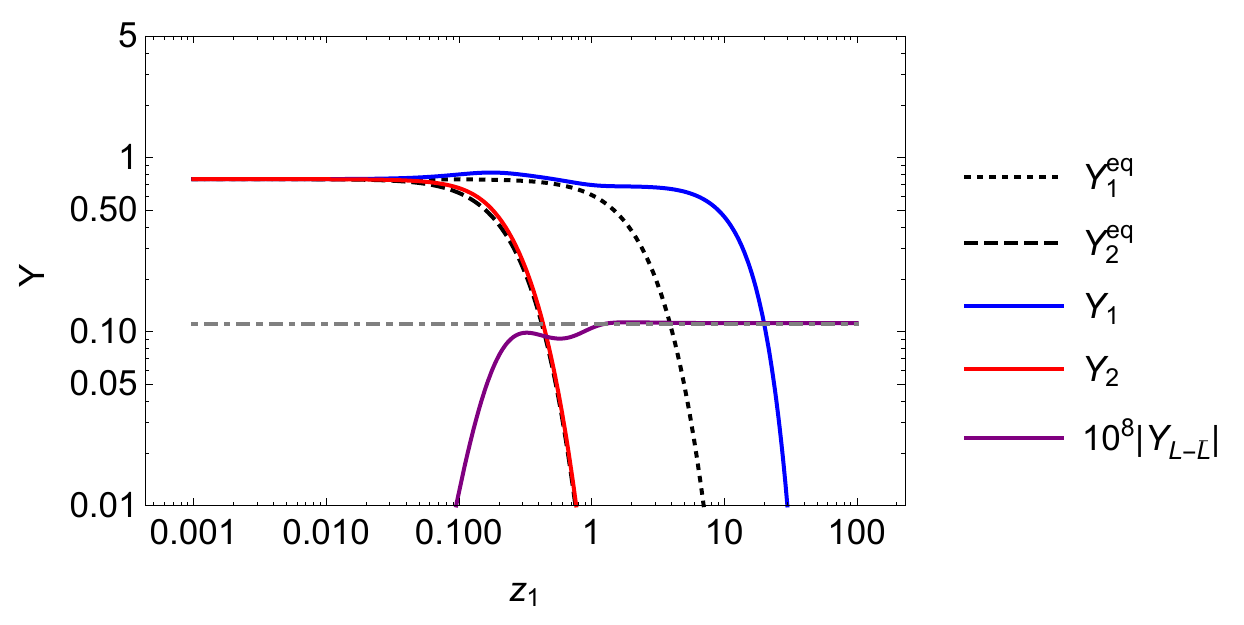}
 	    \caption{ 	\label{fig:benchC} The benchmark scenario (case 1) in Table \ref{tab:bp} with mass hierarchy $M_2=79.5$ TeV and $M_1=8.6$ TeV.
       Shown in the figure are the abundances of the two neutrinos with solid blue and red lines. The dashed lines show
       the corresponding equilibrium densities. The purple curve shows the resulting net lepton abundance normalized by a factor
       $10^{-8}$ so that the value corresponding to observed BAU is $Y_{L-\bar{L}}=0.11\cdot 10^{-8}$ shown by the horizontal
       dot-dashed line.
       }
 	\end{center}
     \end{figure}

    The value of the lepton asymmetry is rather sensitive to the value of $\epsilon$, and the parameter $M_0$ is constrained to be
    relatively small, $M_0\lesssim 3$ TeV (see Fig.~\ref{fig:M0lim}), from the Higgs mass requirement.
    In Fig.~\ref{fig:benchC} we show the
    resulting abundances of the two neutrino species obtained from the Boltzmann equations together with the corresponding equilibrium
    densities, and the net lepton abundance.
    To obtain the observed BAU,  the net lepton abundance must be
    $\eta_L\equiv Y_{L-\bar{L}}=1.1\cdot 10^{-9}$
    The  cross sections of the relevant processes  are given in Appendix~\ref{app:crossSec}.
    A benchmark set of parameters (case 1) leading to the desired value is given by $\epsilon=2.3\cdot 10^{-4}$\,, $|\alpha_0|=1.23\cdot 10^{-3}$
    and $M_0=0.4$ TeV.
    The mass of the heavier neutrino is fixed by the vacuum alignment to 79.5 TeV, while with the above parameters the mass of the lighter one becomes 8.6 TeV.
    The values of the light-neutrino Yukawa couplings, $Y$, enter the analysis via factors $K_1=0.01$, $K_2=4.0$.
    The scalar self-coupling enters through the trilinear
    coupling $\beta\varphi(H^{\dagger}H)$,  $\beta=2.5\cdot 10^{-4}$~TeV.
    Finally, with these parameters the mass of the singlet scalar is $M_\varphi=3.0$~TeV.

    	\begin{table}
	    \caption{We show the values of the model parameters, $K_2$, $\alpha_0$, $\epsilon$, $\beta$, and the values of $v$ and $M_1$ for the cases $1-3$
		(see the text for details). All masses and $v$ are in TeV.
		The heavy neutrino masses, $M_1$ and $M_2$, are determined by the vacuum and the matrix $\alpha$.
		For all the cases we use the same following benchmark values of the remaining parameters:
		$K_1=0.01$, $M_0=0.4$~TeV, and $M_{\varphi}=3.0$~TeV.
		The mass of heavier neutrino is the same, $M_2=79.5$~TeV, for all the cases.
	    }
	    \label{tab:bp}
	    \begin{center}
		\begin{tabular}{cccccccc}
		    \toprule
		    \#$\vphantom{\frac{\frac12}{2}}$
			& \hspace{0.1cm} $K_2$\hspace{0.1cm}
			& \hspace{0.1cm} $10^4\,|\alpha_0|$\hspace{0.1cm}
			&\hspace{0.1cm} $10^4\,\epsilon$\hspace{0.1cm}
			& \hspace{0.1cm} $10^4\,\beta$\hspace{0.1cm}
			&\hspace{.1cm} $10^{-4}v$\hspace{0.1cm} &\hspace{0.1cm} $M_1$\hspace{0.1cm}
			\\
		    \colrule
		    1 $\vphantom{\frac{\frac12}{2}}$  
			& 4.0
			& 12.3	& 2.3 & 2.5 & 3.5 & 8.6 
			\\
		    2 $\vphantom{\frac{\frac12}{2}}$  
			& 4.0
			& 13.5 & 4.6 & 2.5 & 3.5 & 16.7 
			\\
		    3 $\vphantom{\frac{\frac12}{2}}$ 
			& 40
			& 12.3 & 2.3 & 2.5 & 3.5 & 8.6 
			\\
		    \botrule
		\end{tabular}
	    \end{center}
	\end{table}
	\begin{figure}
	    \centering
	    \subfigure[$\ \epsilon\rightarrow 2\epsilon$]{
		\includegraphics[width=0.48\textwidth]{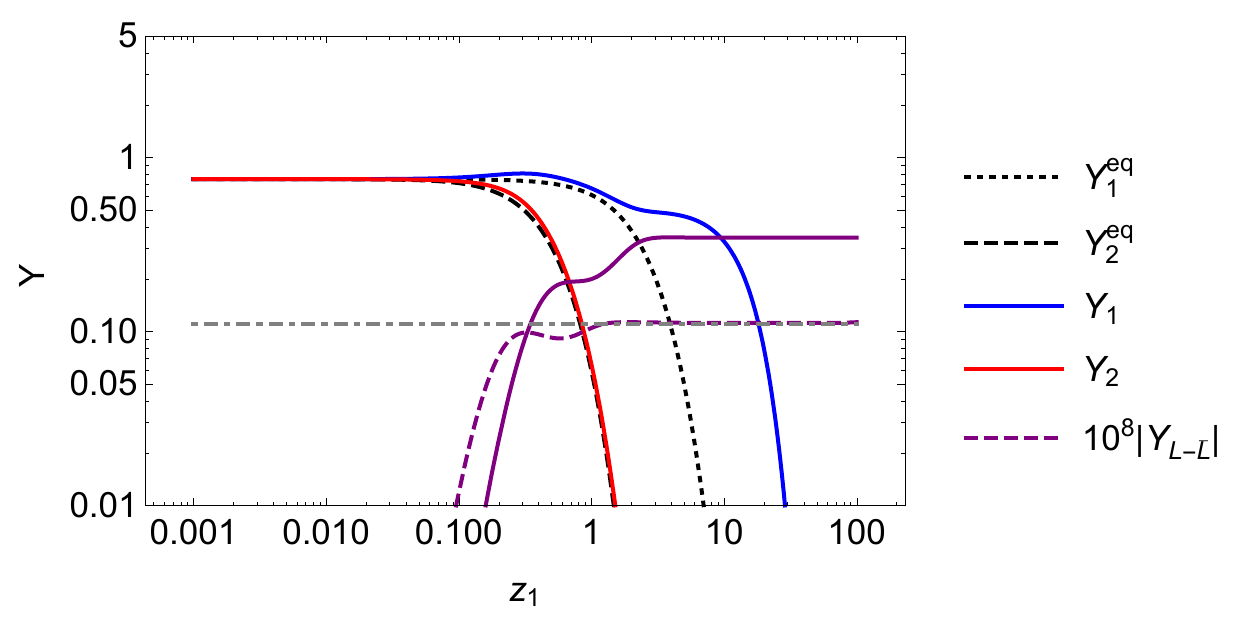}}\\
	    \subfigure[$\ K_2\rightarrow 10K_2$]{\includegraphics[width=0.48\textwidth]{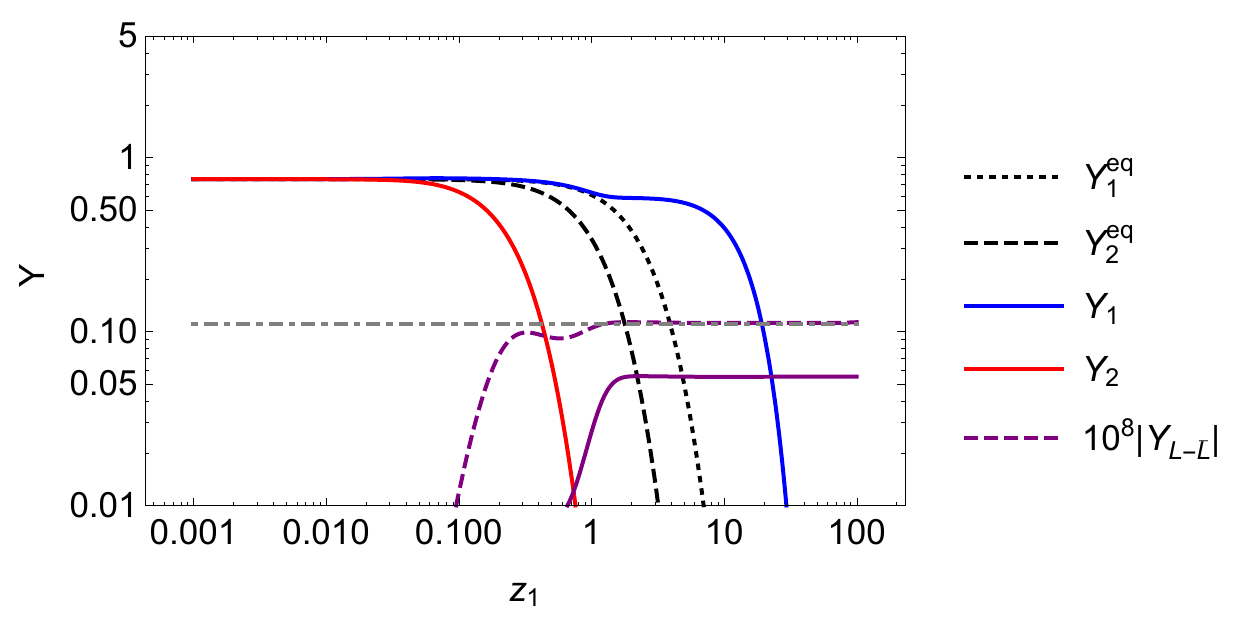}}
	    \caption{We show in this plots a comparison of the main results presented in Fig. \ref{fig:benchC}
		(dashed purple curve) and the cases 2 (upper panel) and 3 (lower panel) presented in Table~\ref{tab:bp}.
		More details in the text.}
	    \label{fig:benchC2}
	\end{figure}

    This result is clearly sensitive to the values of the parameters involved in the analysis. We have checked that changing the parameter $M_0$
    by a factor of two has hardly an
    effect for the asymmetry. This is expected, as $M_0$ does not affect the
    amount of CP violation.  On the other hand, the effects of
    varying $\epsilon$ in Eq.~\eqref{eq:twoCases} and the light neutrino mass parameters entering via $K_2$ have big effects on the
    resulting lepton asymmetry.
    For concreteness, we show in Fig.~\ref{fig:benchC2} the results considering $\epsilon\rightarrow 2 \epsilon$ (case 2),
    and $K_{2}\rightarrow 10 K_{2}$ (case 3).
    In Table \ref{tab:bp} we list the values of the benchmark point (case 1) and the numerical values considered in cases 2 and 3.
    In the second case the final lepton asymmetry is
    enhanced, while in the third case is suppressed. For comparison, the  dashed purple line shows our benchmark point of Fig.~\ref{fig:benchC}.

    From the numerical results we see that the effect due to changing
    $\epsilon$ can be compensated for by changing $K_2$ accordingly. As a
    consequence, one may expect that a sizeable portion of parameter space
    exists, within which the observed BAU can be produced by this model.

\section{Conclusions}
\label{sec:checkout}

In this paper we have considered an extension of the SM where the Higgs boson
arises as a pseudo-Nambu--Goldstone boson.
As a concrete framework,
we have considered the SO(5)/SO(4) coset in terms
of elementary scalar fields.
The scalar content of the theory
is accommodated in the fundamental of SO(5),
and in addition to SM Higgs sector it contains a real
singlet scalar field.
Furthermore, we added two heavy RH neutrinos
to explain the masses and mixings of the light neutrinos
and also to generate BAU via leptogenesis.
The neutrino Majorana masses are generated through the
interplay between a bare mass term and a
mass term arising dynamically
from the vacuum expectation value of the scalar singlet.

We found that the vacuum structure and the mass of the Higgs boson imply
stringent bounds on the model parameters. In particular, we demonstrated
that these lead to the requirement that the
heavier of the two RH neutrino mass eigenvalues is $\sim 80$ TeV.

This in turn has consequences for the generation of the BAU via leptogenesis.
As we have discussed, this depends sensitively on the assumed mass patterns
of the RH neutrinos. We analysed in detail the cases of degenerate and
hierarchical RH neutrinos, and identified examples of viable values
of the parameters.

One could extend the model by adding a third RH neutrino as well.
This would be a suitable candidate for dark matter given a vanishing
mixing of this state with the two other RH neutrinos. The vacuum
analysis would not be affected by its presence as we have shown that
the vacuum alignment is determined only by the heaviest RH neutrino
states. Furthermore, our analysis of BAU would also remain unaffected.

\section*{Acknowledgements}
This work has been supported by the Academy of Finland, grant\# 267842. The CP$^3$-Origins centre is partially funded
by the Danish National Research Foundation, grant number DNRF90. TA acknowledges partial funding from a Villum foundation grant.

\onecolumngrid
\clearpage

\appendix

\section{Definitions }

\subsection{Definitions for resonant leptogenesis}
\label{app:resonant}

To estimate the $B-L$ asymmetry, we follow Refs.~\cite{Pascoli:2006ie,Pascoli:2006ci,DiBari:2016guw}. The final asymmetry is
given as a sum of the contributions from the two lightest RH neutrinos,
and three lepton flavours
\begin{equation}
    \label{eq:nbminusl}
	N_{B-L}^{\mathrm{f}}=\sum_{i, \alpha}N_{\Delta_{\alpha}}^{i}
	    =\sum_{i, \alpha}\epsilon_{i\alpha}\kappa(K_{1\alpha}+K_{2\alpha}),
\end{equation}
where $i=1,2$, $\alpha=e,\mu,\tau$, and we  define:
\begin{equation}
    \label{eq:}
    \epsilon_{i\alpha}\equiv-\frac{\Gamma_{i\alpha}-\bar{\Gamma}_{i\alpha}}{\sum_{\alpha}\left(\Gamma_{i\alpha}
	+\bar{\Gamma}_{i\alpha}\right)}, \quad
    K_{i\alpha}\equiv\frac{\Gamma_{i\alpha}+\bar{\Gamma}_{i\alpha}}{H(T=M_i)}=\frac{\left|m_{D\,\alpha\,i}\right|^2}{M_im_{\star}},
\end{equation}
and
\begin{equation}
    \label{eq:}
    \kappa(x)\equiv\frac{2}{x\,z_B(x)}\left[1-\exp\left(-\frac{x\,z_B(x)}{2}\right)\right],
\quad \mbox{where} \quad
    z_B(x)\simeq 2+4x^{0.13}\ee^{-\frac{2.5}{x}}.
\end{equation}
In the degenerate limit, the final asymmetry can be written as
\begin{equation}
    \label{eq:}
    \begin{split}
	N_{B-L}^{\mathrm{f}}\simeq&\frac{\bar{\epsilon}(M_1)}{3\dl}\left(\frac{1}{\bar{K}_1}+\frac{1}{\bar{K}_2}\right) \sum_{\alpha}\kappa(K_{1\alpha}
	    +K_{2\alpha})\left(\mathcal{I}^{\alpha}_{12}
	    +\mathcal{J}^{\alpha}_{12}\right),
    \end{split}
\end{equation}
where
\begin{equation}
    \label{eq:}
    \begin{split}
	\mathcal{I}^{\alpha}_{ij}&\equiv\frac{\mathrm{Im}\left[m_{D\,\alpha\,i}^*m_{D\,\alpha\,j}
	    (m_{D}^{\dagger}m_{D})_{ij}\right]}{M_iM_jm_{\mathrm{atm}}m_{\star}},\\
	\mathcal{J}^{\alpha}_{ij}&\equiv\frac{\mathrm{Im}\left[m_{D\,\alpha\,i}^*m_{D\,\alpha\,j}
	    (m_{D}^{\dagger}m_{D})_{ji}\right]}{M_iM_jm_{\mathrm{atm}}m_{\star}}\frac{M_i}{M_j},
    \end{split}
\end{equation}
and
\begin{equation}
    \bar{K}_i\equiv\sum_{\alpha}K_{i\alpha},\quad i=1,2.
\end{equation}

Assuming that the baryon-to-photon ratio at recombination satisfies $\eta_B\simeq 0.01 \,N_{B_L}^{f}$, for normal hierarchy
it is found
\begin{equation}
    \label{eq:}
    \begin{split}
    \dl\simeq &\,0.01\frac{\bar{\epsilon}(M_1)}{\eta_B}f(m_{\nu},\Omega)\\
    \simeq&\, 0.8\times 10^{-5}\left(\frac{f(m_{\nu},\Omega)}{f_{\max}}\right)\left(\frac{M_1}{10^6\,\mathrm{GeV}}\right),
    \end{split}
\end{equation}
where $f_{\max}\simeq 0.005$, and the function $f$ incorporates the dependence on the light-neutrino masses and mixings:
\begin{equation}
    \label{eq:}
    \begin{split}
	f(m_{\nu},\Omega)\equiv& \frac{1}{3}\left(\frac{1}{\bar{K}_1}+\frac{1}{\bar{K}_2}\right)\sum_{\alpha}\kappa(K_{1\alpha}+K_{2\alpha})\left(\mathcal{I}^{\alpha}_{12}
	    +\mathcal{J}^{\alpha}_{12}\right).
    \end{split}
\end{equation}

\subsection{Definitions for the  hierarchical case}
\label{app:def}

In this Appendix we give the decay, scattering and washout functions that appear in the Boltzmann equations, Eq.~\eqref{eq:BEs}. We follow the notations
of Ref.~\cite{Dall:2014nma}. We define
\begin{equation}
    \label{eq:}
    W_{1D_i}\equiv\frac{1}{2Y_L^{\mathrm{eq}}}D_i\quad \mbox{where}\quad   D_i\equiv\frac{\gamma^{\mathrm{eq}}_{D_i}}{n_{\gamma}^{\mathrm{eq}}H}
	=K_iz_i^2Y_i^{\mathrm{eq}}\frac{K_1(z_i)}{K_2(z_i)},
\end{equation}
\begin{equation}
    \label{eq:}
    D_{21}\equiv K_{21}z_2^2Y_2^{\mathrm{eq}}\frac{K_1(z_2)}{K_2(z_2)}.
\end{equation}
The scattering function for the process $i j \rightarrow mn$ reads:

\begin{equation}
    \label{eq:}
    S_{ij\rightarrow mn}\equiv\frac{\gamma^{\mathrm{eq}}_{ij\rightarrow mn}}{n_{\gamma}^{\mathrm{eq}}H}=\frac{m_i}{g_{\gamma}H_i}\frac{1}{32\pi^2}z_i
    \int_{w_{\min}}^{\infty}\mathrm{d}w\sqrt{w}K_1(\sqrt{w})\hat{\sigma}_{ij\rightarrow mn}(w\frac{m_i^2}{z_i^2}),
\end{equation}
\begin{equation}
    \label{eq:}
    w_{\min}=\max\{(m_i+m_j)^2,(m_m+m_n)^2\}.
\end{equation}
We define
\begin{equation}
    \label{eq:}
    \begin{split}
	W_{S_1}=&\frac{1}{Y_L^{\mathrm{eq}}}\left(2S_{N_1t\rightarrow LQ}+S_{N_1H\rightarrow L\varphi}+S_{N_1\varphi\overset{1}{\rightarrow} LH}\right)
	    +\frac{Y_1}{Y_L^{\mathrm{eq}}Y_1^{\mathrm{eq}}}\left(S_{N_1L\rightarrow Qt}+S_{N_1L\rightarrow H\varphi}\right),\\
	W_{S_2}=&\frac{1}{Y_L^{\mathrm{eq}}}\left(2S_{N_2t\rightarrow LQ}+S_{N_2H\rightarrow L\varphi}+S_{N_2S\overset{1}{\rightarrow} LH}
	    +S_{N_2S\overset{2}{\rightarrow} LH}\right)
	    +\frac{Y_2}{Y_L^{\mathrm{eq}}Y_2^{\mathrm{eq}}}\left(S_{N_2L\rightarrow Qt}+S_{N_2L\rightarrow H\varphi}\right),
    \end{split}
\end{equation}
and
\begin{equation}
\label{eq:}
    \begin{split}
	S_1=&2S_{N_1L\rightarrow Qt}+4S_{N_1Q\rightarrow Lt}+2S_{N_1L\rightarrow H\varphi}+4S_{N_1H\rightarrow L\varphi}
	    +2S_{N_1\varphi\overset{1}{\rightarrow} LH}\\
	S_2=&2S_{N_2L\rightarrow Qt}+4S_{N_2Q\rightarrow Lt}+2S_{N_2L\rightarrow H\varphi}+4S_{N_2H\rightarrow L\varphi}
	    +2S_{N_2\varphi\overset{1}{\rightarrow} LH}+2S_{N_2\varphi\overset{2}{\rightarrow} LH}.
    \end{split}
\end{equation}

\section{Cross sections}
\label{app:crossSec}

All the cross sections are summed over \emph{both} the initial and final state degrees of freedoms (gauge, spins, lepton flavour). In some of the formulas
we have found slightly different coefficients compared to \cite{Dall:2014nma}.

    \subsection{Decay widths}
    \label{app:decayW}
	\begin{equation}
	    \label{eq:}
	    \Gamma_i\equiv \Gamma(N_i\rightarrow LH)+\Gamma(N_i\rightarrow \overline{LH})=\frac{(YY^{\dagger})_{ii}M_i}{8\pi}.
	\end{equation}
	\begin{equation}
	    \label{eq:gammaij}
		\Gamma_{ji}\equiv \Gamma(N_j\rightarrow N_i\varphi)
		=\frac{|(\alpha^\dagger\alpha)_{ji}|^2M_j}{16\pi}\left[\left(1+\frac{M_i}{M_j}\right)^2
		-\frac{\MS^2}{M_j^2}\right]
		\cdot\sqrt{\left(1-\frac{M_i^2}{M_j^2}-\frac{\MS^2}{M_j^2}\right)^2-4\frac{M_i^2}{M_j^2}\frac{\MS^2}{M_j^2}}.
	\end{equation}

    \subsection{Scattering cross sections}
    \label{app:scatt}

    Below we tabulate the relevant cross sections $\sigma(ij\rightarrow mn)$. The reduced cross sections are obtained from these as
    \begin{equation}
	\label{eq:}
	\hat{\sigma}(ij\rightarrow mn)=\frac{1}{s}\delta(s,m_i^2,m_j^2)\sigma(ij\rightarrow mn),
    \end{equation}
    where
    \begin{equation}
	\label{eq:}
	\delta(a,b,c)=a^2+b^2+c^2-2ab-2ac-2bc.
    \end{equation}

\begin{enumerate}[a.]

    \item  $N_iL\rightarrow Q\bar{t}$
	\begin{flalign}
	    \label{eq:}
	    &\sigma(N_iL\rightarrow Q\bar{t})=\frac{N_c\,(YY^{\dagger})_{ii}\,y_t^2}{8\pi}\frac{s}{(s-m_h^2)^2}.&&
	\end{flalign}
	\vspace{0.4cm}
        \item $N_iQ\rightarrow Lt$
	\begin{flalign}
	    \label{eq:}
	    &\sigma(N_iQ\rightarrow Lt)=\frac{N_c\,(YY^{\dagger})_{ii}\,y_t^2}{8\pi(s-M_i)^2}\left[\frac{s-2M_i^2+2m_h^2}{s-M_i^2+m_h^2}
		+\frac{M_i^2-2m_h^2}{s-M_i^2}\log\frac{s-M_i^2+m_h^2}{m_h^2}\right].&&
	\end{flalign}
	\vspace{0.4cm}
        \item $N_iL\rightarrow H\varphi$
	\begin{flalign}
	    \label{eq:}
	    &\sigma(N_iL\rightarrow H\varphi)=\frac{(YY^{\dagger})_{ii}\beta^2}{8\pi}\frac{s-\MS^2}{s^3}.&&
	\end{flalign}
	\vspace{0.4cm}
	\item $N_i\varphi\rightarrow LH$ via $N_j$
	\begin{flalign}
	    \label{eq:}
	    &\sigma(N_i\varphi\overset{j}{\rightarrow} LH)=\frac{(YY^{\dagger})_{ii}|(\alpha \alpha^{\dagger})_{ji}|^2}{16\pi(s-M_j^2)^2
		\sqrt{\delta(s,M_i^2,\MS^2)}}\left[(s+M_i^2-\MS^2)(s+M_j^2)-4M_iM_j s\right].&&
	\end{flalign}
	\vspace{0.4cm}
	\item $N_iN_j\rightarrow HH$
	\begin{flalign}
	    \label{eq:}
	    &\sigma(N_iN_j\rightarrow HH)=\frac{|(\alpha \alpha^{\dagger})_{ji}|^2\beta^2}{8\pi}\frac{s-(M_i+M_j)^2}{(s-\MS^2)^2
		\sqrt{\delta(s,M_i^2,M_j^2)}}.&&
	\end{flalign}
	\vspace{0.4cm}
	\item $N_iH\rightarrow L\varphi$
	\begin{flalign}
	    \label{eq:}
	    &\sigma(N_iH\rightarrow L\varphi)=\frac{(YY^{\dagger})_{ii}\beta^2}{16\pi (s-M_i^2)^2}\log\frac{s^2(s-M_i^2-\MS^2)^2
		+s^2\mathcal{E}_i^2}{M_i^4\MS^4+s^2\mathcal{E}_i^2}.&&
	\end{flalign}

    \end{enumerate}

\clearpage
\twocolumngrid

\bibliography{References}

\end{document}